\documentclass{appolb}
\usepackage{epsfig}

\begin{document}
\title{ Numerical evaluation of some master integrals \\
for the 2-loop general massive self-mass \\
from differential equations.
\thanks{ Presented at {\sl Matter To The Deepest,
 XXVII ICTP, Ustron, 15-21 Sept 2003 }, \\
Supported in part by the EC network EURIDICE, 
contract HPRN-CT2002-00311.}
}
\author{Michele Caffo
\address{ {\sl INFN, Sezione di Bologna, I-40126 Bologna, Italy} and \\
{\sl Dipartimento di Fisica, Universit\`a di Bologna, I-40126 Bologna, Italy} 
} }
\maketitle
  
\def\Re{\hbox{Re~}} 
\def\Im{\hbox{Im~}} 
\newcommand{\Eq}[1]{Eq.(\ref{#1})} 
\newcommand{\Pol}[2]{P_{#1,#2}(m_i^2,p^2)} 
\newcommand{\Qol}[2]{Q_{#1,#2}(m_i^2,p^2)} 
\newcommand{\D}{D(m_i^2,p^2)} 
\newcommand{\F}[1]{F_{#1}(n,m_i^2,p^2)} 
\newcommand{\Fn}[2]{F_{#1}^{(#2)}(m_i^2,p^2)} 
\newcommand{\Sm}{S(n,m_1^2,m_2^2,p^2)} 
\newcommand{\dnk}[1]{ \frac{d^nk_{#1}}{(2\pi)^{n-2}} }
\begin{abstract}
The 4-th order Runge-Kutta method in the complex plane is proposed 
for numerically advancing the solutions of a system of first order 
differential equations in one external invariant satisfied by the 
master integrals related to a Feynman graph.
Some results obtained for the 2-loop self-mass MI are reviewed.
The method offers a reliable and robust approach to the direct and precise 
numerical evaluation of master integrals.
\end{abstract}
\PACS{\\ 
PACS 11.10.-z Field theory \\ 
PACS 11.10.Kk Field theories in dimensions other than four \\ 
PACS 11.15.Bt General properties of perturbation theory    \\ 
PACS 12.20.Ds Specific calculations  \\
PACS 12.38.Bx Perturbative calculations
}

\section{Introduction}

The aim to more precise and handy calculations of radiative corrections 
push for a restless research and development of new methods. 

Today the organization of the calculations is usually based on the 
integration by part identities and on the evaluation of the master integrals 
(MI) \cite{TkaChet}.
In this frame the differential equations for the MI, or Master Differential 
Equations (MDE), can be used not only for their analytic calculations 
when the number of the variables and parameters is small, but also for their 
direct numerical evaluation, when the large number prevents the success of 
an analytic calculation, in alternative to the more commonly used integration 
methods or to the more recently introduced difference equations method. 

Here is presented a method to get a numerical solution of the MI for any
values of the variables and parameters, using directly the MDE and the 
values of the MI known in a given set of the variables and parameters. 
Some results obtained for the 2-loop self-mass MI are reviewed.

\section{Master Differential Equations}

Starting from the integral representation of the $N_{MI}$ MI, 
related to a certain Feynman graph, 
by derivation with respect to one of the internal masses $m_i$ \cite{Kotikov} 
or one of the external invariants $s_e$ \cite{Remiddi} and with the repeated 
use of the integration by part identities, a system of $N_{MI}$ independent 
first order partial MDE is obtained for the $N_{MI}$ MI. 
For any of the $s_e$, say $s_j$, the equations have in general the form 
\begin{eqnarray} 
&&K_k(m_i^2,s_e)
\ {\frac{\partial}{\partial s_j}} F_k(n,m_i^2,s_e) = \nonumber\\
&&{\kern-20pt}\sum_{l} M_{k,l}(n,m_i^2,s_e) F_l(n,m_i^2,s_e)
+T_k(n,m_i^2,s_e), \nonumber\\ 
&& k,l=1,...,N_{MI} \label{mde}
 \end{eqnarray} 
where $F_k(n,m_i^2,s_e)$ are the MI,  
$K_k(m_i^2,s_e)$ and $M_{k,l}(n,m_i^2,s_e)$ are polynomials, 
while $T_k(n,m_i^2,s_e)$ are polynomials times simpler MI of the 
subgraphs of the considered graph.
The roots of the equations 
\begin{equation} 
K_k(m_i^2,s_e) = 0 
\label{sp}
\end{equation} 
identify the {\em special} points, where numerical calculations are 
troublesome.  
Fortunately analytic calculations at those points come out to be 
possible in all the attempted cases so far.
They might not be simple and often require some external knowledge, 
like the assumption of regularity of the solution at that {\em special} point. 

To solve the system of equations it is necessary to know the MI for a 
chosen value of the differential variable, $s_j$ in \Eq{mde}.
For that purpose we use the analytic expressions at the {\em special} 
points, taken as the starting points of the advancing solution path. 
Moreover starting from one {\em special} point, not only the values of 
the MI are necessary, but also their first order derivatives at that point. 
That is because some of the coefficients $K_k(m_i^2,s_e)$ of the MI 
derivatives in the differential equations \Eq{mde} vanish at that point. 
Therefore also the analytic expressions for the first derivatives of MI
at {\em special} points are obtained, but this usually comes out to be a 
simpler task (unless poles in the limit of the number of dimensions $n$ 
going to 4 are present). 

Enlarging the number of loops and legs increases the number of parameters, 
MI and equations, but does not change or spoil the method. 

\section{The 4-th order Runge-Kutta method}

Many methods are available for obtaining the numerical solutions of 
a first-order differential equation \cite{RK} 
\begin{equation} 
\frac{\partial y(x)}{\partial x} = f(x,y) \ .
\label{fode}
\end{equation}
The 4-th order Runge-Kutta method is a rather precise and robust approach 
to advance the solution from a point $x_n$, where the solution 
$y_n$ is known, to the point $x_{n+1} = x_n + \Delta$ 
By suitably choosing the intermediate points where calculating $f(x,y)$ 
one obtains the 4-th order Runge-Kutta formula 
\begin{eqnarray} 
 k_1 &=& \Delta f(x_n,y_n), \ \nonumber \\
 k_2 &=& \Delta f(x_n+\frac{\Delta}{2},y_n+\frac{k_1}{2}), \ \nonumber \\
 k_3 &=& \Delta f(x_n+\frac{\Delta}{2},y_n+\frac{k_2}{2}), \ \nonumber \\
 k_4 &=& \Delta f(x_n+\Delta,y_n+k_3), \ \nonumber \\
 y_{n+1} &=& y_n +\frac{k_1}{6}+\frac{k_2}{3}+\frac{k_3}{3}+\frac{k_4}{6} 
+{\cal O}(\Delta^5) 
\label{4ork}
\end{eqnarray} 
which omits terms of order $\Delta^5$.

To avoid numerical problems due to the presence of {\em special} points 
on the real axis, it is convenient to choose a path for advancing the 
solution in the complex plane of $x$.

The extension from one first-order differential equation to a system of 
$N_{MI}$ first-order MDE for the $N_{MI}$ MI is straightforward \cite{RK}. 

\section{Results}

I report here some results obtained with the outlined method.

\begin{figure}[htb]
\vspace{9pt}
{\scalebox{.7}[.7]{\includegraphics*[80,20][370,120]{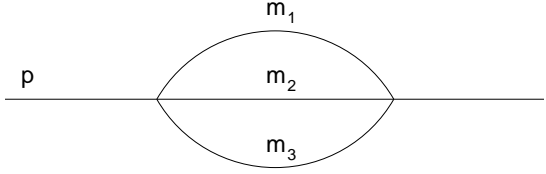}}}
\caption{The general massive 2-loop sunrise self-mass diagram.}
\label{fig:sunrise}
\end{figure}

\begin{figure}[htb]
{\scalebox{.7}[.7]{\includegraphics*[20,40][260,140]{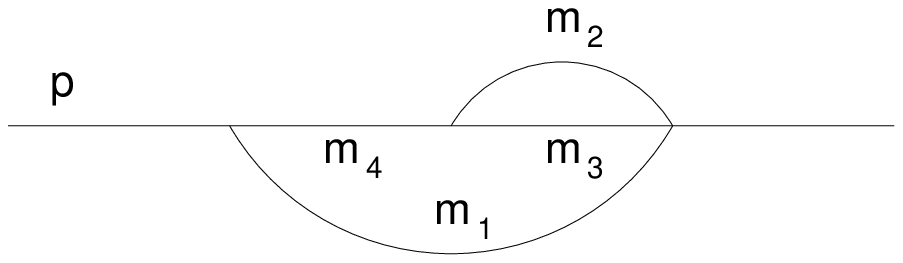}}}
\caption{The general massive 2-loop 4-denominator self-mass diagram.}
\label{fig:4den}
\end{figure}

\begin{figure}[htb]
{\scalebox{.7}[.7]{\includegraphics*[20,20][260,120]{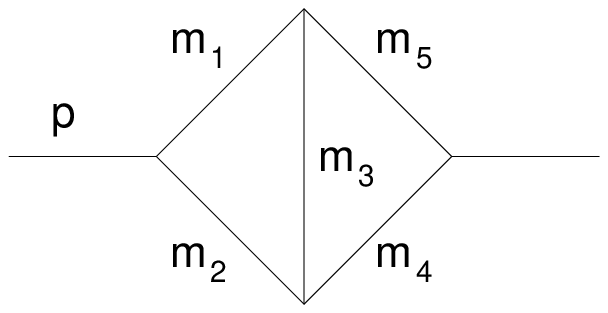}}}
\caption{The general massive 2-loop 5-denominator self-mass diagram.}
\label{fig:5den}
\end{figure}

The simplest nontrivial application of the method is to the 2-loop 
self-mass with arbitrary internal masses, which has three 2-loop 
topologies \cite{Tarasov}: 
the sunrise, shown in Fig.\ref{fig:sunrise}, with 4 MI,
the one with 4 propagators, shown in Fig.\ref{fig:4den}, with one more MI, 
and the one with 5 propagators, shown in Fig.\ref{fig:5den}, with 
one even more MI. 
The simpler 2-loop self-mass amplitudes of 
Fig.\ref{fig:sunrise},\ref{fig:4den},\ref{fig:5den}
can be written in the integral form as 
\begin{eqnarray} 
 && A(n,m_i^2,p^2,-\alpha_1,-\alpha_2,-\alpha_3,-\alpha_4,-\alpha_5) 
     = \mu^{8-2n} \nonumber \\  
 && \int \dnk{1} \int \dnk{2} \; 
    \frac{ 1 } { D_1^{\alpha_1} D_2^{\alpha_2} D_3^{\alpha_3} 
                                D_4^{\alpha_4} D_5^{\alpha_5} }, \nonumber \\
 && D_i=(k_i^2+m_i^2), \qquad i=1,2,3,4,5.
\end{eqnarray} 
where $k_i$ is the 4-momentum of the line of mass $m_i$ and 4-momentum 
conservation is understood in the vertexes.
The arbitrary mass scale $\mu$ accounts for the continuous value 
of the dimensions $n$; as one of the natural scales of the problem 
is the 3-body threshold of the sunrise amplitudes, we take usually 
$\mu = m_1 + m_2 +m_3$. 

The 4 MI of the sunrise, the one of the 4-denominator and the one of the 
5-denominator can be chosen as 
\begin{eqnarray} 
  \F{0}&=&A(n,m_i^2,p^2,-1,-1,-1,0,0) \ ,\nonumber \\ 
  \F{1}&=&A(n,m_i^2,p^2,-2,-1,-1,0,0) \ ,\nonumber \\ 
  \F{2}&=&A(n,m_i^2,p^2,-1,-2,-1,0,0) \ ,\nonumber \\ 
  \F{3}&=&A(n,m_i^2,p^2,-1,-1,-2,0,0) \ ,\nonumber \\ 
  \F{4}&=&A(n,m_i^2,p^2,-1,-1,-1,-1,0) \ ,\nonumber \\ 
  \F{5}&=&A(n,m_i^2,p^2,-1,-1,-1,-1,-1) \ , 
\end{eqnarray} 
the other amplitudes are related to these by integration by parts identities 
\cite{Tarasov,MS,CCLR1}, which are used also to obtain the system of MDE 
for the MI in the form of \Eq{mde}. 

In the MDE of the sunrise MI the only lower order MI entering is the one 
related to the 1-loop vacuum graph
\begin{eqnarray}
 T(n,m^2) = \int \dnk{} \frac{1}{k^2+m^2} = \frac{m^{n-2}C(n)}{(n-2)(n-4)}\ , 
\end{eqnarray}
while for the 4-denominator MDE is necessary also the knowledge of the 1-loop 
self-mass and the 2-loop vacuum MI, known analytically, and the sunrise MI. 
The 5-denominator MDE requires also the knowledge of the 4-denominator MI.

The function
$C(n) = \left(2 \sqrt{\pi} \right)^{(4-n)} \Gamma\left(3-\frac{n}{2}\right)$, 
which appears in the expressions for the MI as an overall factor with an 
exponent equal to the number of loops, is usually 
kept unexpanded in the limit $n \to 4$, and only at the very end of the 
calculation for finite quantities is set $C(4) = 1$. 

When the MI are expanded in $(n-4)$, for $k=0,1,2,3,4,5$ and 
$i=1,2,3,4,5$
\begin{eqnarray} 
\F{k} = C^2(n) && \Biggl\{ \frac{1}{(n-4)^2} \Fn{k}{-2} \nonumber \\
 + \frac{1}{(n-4)} \Fn{k}{-1} &+& \Fn{k}{0} 
 + {\cal O} (n-4) \Bigr\} \ , 
\end{eqnarray} 
the coefficients of the poles can be easily obtained analytically for 
arbitrary values of the external squared momentum $p^2$, 
\begin{eqnarray} 
 \Fn{0}{-2} &=& -\frac{1}{8} (m_1^2+m_2^2+m_3^2) \ ,
\nonumber \\
 \Fn{0}{-1} &=&  \frac{1}{8} \Biggl\{ \frac{p^2}{4} 
                +\frac{3}{2} (m_1^2+m_2^2+m_3^2) \nonumber \\  
&& {\kern-90pt}
 - \left[ m_1^2 \log\left(\frac{m_1^2}{\mu^2}\right) 
 +m_2^2 \log\left(\frac{m_2^2}{\mu^2}\right)
 +m_3^2 \log\left(\frac{m_3^2}{\mu^2}\right) \right] 
                        \Biggr\} \ , \nonumber \\
 \Fn{k}{-2} &=& \frac{1}{8} \ , \quad k=1,2,3 \nonumber \\
 \Fn{k}{-1} &=&  - \frac{1}{16} 
 + \frac{1}{8} \log\left(\frac{m_k^2}{\mu^2}\right) \ , \nonumber \\
 \Fn{4}{-2} &=& + \frac{1}{8} \ , \nonumber \\ 
 \Fn{4}{-1} &=& - \frac{1}{16}  
 - \frac{1}{2} S^{(0)}(m_1^2,m_4^2,p^2) \ ,
\end{eqnarray} 
where $S^{(0)}(m_1^2,m_4^2,p^2)$ is the finite part of the $(n-4)$ 
expansion of the 1-loop self-mass
\begin{equation} 
  \Sm = C(n) \Biggl\{ - \frac{1}{2} \frac{1}{(n-4)} 
       + S^{(0)}(m_1^2,m_2^2,p^2) + {\cal O} (n-4)\Biggr\} \ ,
\end{equation} 
which is known analytically. The 5-denominator is not divergent for
$n\simeq4$, so $\Fn{5}{-2} = \Fn{5}{-1} = 0$ and $\F{5} = \Fn{5}{0}$.
The finite parts $\Fn{k}{0}$ satisfy MDE of the type of \Eq{mde}.

The {\em special} points are easily obtained from the coefficient 
$K_k(m_i^2,p^2)$. For the sunrise MI they are at $p^2=0, \infty$ and 
the roots of 
\begin{eqnarray} 
\D  &=& \left[p^2+(m_1+m_2+m_3)^2\right] 
        \left[p^2+(m_1+m_2-m_3)^2\right] \nonumber \\
    &&{\kern-20pt}  \left[p^2+(m_1-m_2+m_3)^2\right] 
        \left[p^2+(m_1-m_2-m_3)^2\right]  = 0 \ , 
\end{eqnarray}
For the sunrise MI the analytic expressions for their first order
expansion were completed around the {\em special} points 
\cite{CCLR1,CCR1,CCR2,CCR3}: 
$p^2=0$;  $p^2=\infty$;  $p^2=-(m_1+m_2+m_3)^2$, 
the threshold; $p^2=-(m_1+m_2-m_3)^2$, the pseudo-thresholds.
For the 4-denominator are {\em special} points also the 2-body 
threshold $p^2=-(m_1+m_4)^2$ and pseudo-threshold $p^2=-(m_1-m_4)^2$;
the expansion at $p^2=0$ is completed in \cite{Martin} and 
\cite{CCLR2,CCGR}. 
The 5-denominator has {\em special} points at $p^2=0$, $p^2=\infty$ 
and in some mass combinations of 2-body and 3-body thresholds and 
pseudo-thresholds.

To obtain numerical results for arbitrary values of $p^2$, a 
4th-order Runge-Kutta formula is implemented in a computer code, 
with a solution advancing path starting from the {\em special} points, 
so that also the first term in the expansion is necessary.

The path followed starts from $p^2=0$ and moves in the lower 
half complex plane of $p_r^2 \equiv p^2/\mu^2$, to avoid proximity 
to the other {\em special} points, which can cause loss in precision.
Values between {\em special} points can be safely reached 
through a complex path. 
Results for arbitrary values of $p_r^2$ and of the masses are obtained 
with this method for the sunrise MI in \cite{CCLR1,CCR3} and \cite{Martin}, 
for the 4-denominator MI in \cite{Martin} and \cite{CCLR2,CCGR}, 
for the 5-denominator MI in \cite{Martin}. 
 
For values of $p_r^2$ very close to a {\em special} point 
($\simeq 10^{-4}$) the method fails. 
In the case of the sunrise MI we start the Runge-Kutta path from 
the analytical expansion at that {\em special} point \cite{CCR3}.
A simpler and faster possibility (proposed also in \cite{Martin}) 
is to fit an approximant (the expansion around that {\em special} 
point up to the requested precision) for values of $p^2$ where the 
method works, then using it closer to the {\em special} point. 
A test performed in the case of the sunrise MI where we can check the 
results show that the method works rather well with some cautions 
\cite{CCGR}. 
 
Subtracted differential equations are used when starting from $p^2=\infty$ 
or from threshold, as that points are not regular points of the MDE.

Remarkable self-consistency checks are easily provided by comparing 
the results obtained either starting from the same point and choosing 
different paths to arrive to the same final point, or choosing directly 
different starting points and again arriving to the same final point.

The execution of the program is rather fast and precise: with an 
Intel Pentium III of 1 GHz we get values with 7 digits requiring 
times ranging from a fraction of a second to 10 seconds of CPU, 
and with 11 digits from few tens of seconds to one hour. 

If $\Delta=L/N$ is the length of one step, $L$ is the length of the 
whole path and $N$ the total number of steps, the 4th-order 
Runge-Kutta formula discards terms of order $\Delta^5$, so the 
whole error behaves as $\epsilon_{RK} = N \Delta^5 = L^5/N^4$, and a proper 
choice of $L$ and $N$ allows the control of the precision. 

Indeed we estimate the relative error, as usual, by comparing a value 
obtained with $N$ steps with the one obtained with $N/10$ steps, 
$\epsilon_{RK} = [V(N)-V(N/10)]/V(N)$, to which we add a cumulative 
rounding error $\epsilon_{cre} = \sqrt{N} \times 10^{-15}$, 
due to our 15 digits double precision FORTRAN implementation. 

The general massive sunrise MI are numerically well studied in literature 
and several numerical methods are developed, such as multiple expansions 
\cite{BBBS}, or numerical integration \cite{BBBS,BBBB,GvdB,PT,GKP,P}.
Comparisons are presented in \cite{CCR3} with some values available in the 
literature \cite{BBBS,P} with excellent agreement (up to more 
than 11 digits).
For the 4-denominator MI in \cite{CCGR} we obtain complete agreement with 
\cite{Martin} and with \cite{BT,BBBB}. Calculations via numerical integration 
are also in \cite{PU}.
 
\section{Perspectives}

The presented method for numerically advancing the solutions of the MDE 
is rather precise and competitive with other available methods 
for numerical MI calculations.

It seems to be possible to complete the 2-loop self-mass 
for arbitrary internal masses and we have completed the 4-denominators
case \cite{CCGR}. 

We think that the extension to graphs with more loops or legs do not 
present serious problems, even if the growth in the number of MI increases 
the computing time. 

It is worth to mention that the method relies on the same MDE, which are 
used also for analytic calculations, so it provides a 'low-cost' comforting 
cross-check for those results. 
\def\NP{{\sl Nucl. Phys.}} 
\def\PL{{\sl Phys. Lett.}} 
\def\PR{{\sl Phys. Rev.}} 
\def\PRL{{\sl Phys. Rev. Lett.}} 
\def\NC{{\sl Nuovo Cim.}}
\def\APP{{\sl Acta Phys. Pol.}}
\def\ZP{{\sl Z. Phys.}}
\def\MPL{{\sl Mod. Phys. Lett.}} 
\def\EPJ{{\sl Eur. Phys. J.}} 
\def\IJMP{{\sl Int. J. Mod. Phys.}} 
\def\CPC{{\sl Comput. Phys. Commun.}}

\end{document}